\documentclass[twocolumn,trackchanges]{aastex62}

\pdfoutput=1 
\usepackage{amsmath,amstext}
\usepackage[T1]{fontenc}
\usepackage{apjfonts} 
\usepackage[figure,figure*]{hypcap}

\newcommand{\fracbrac}[2]{\left(\frac{#1}{#2}\right)}
\newcommand{\pd}[2]{\frac{\partial #1}{\partial #2}}

\newcommand{\gsim}{\lower.7ex\hbox{$\;\stackrel{\textstyle>}{\sim}\;$}}
\newcommand{\lsim}{\lower.7ex\hbox{$\;\stackrel{\textstyle<}{\sim}\;$}}
\newcommand{\Z}{\cal Z}
\newcommand{\ReZ}{\text{Re}[{\cal Z}]}
\newcommand{\ImZ}{\text{Im}[{\cal Z}]}

\newcommand{\osc}[1]{\left\{#1\right\}_\text{osc.}} 
\newcommand{\Kepler}{{\it Kepler~}}

\shorttitle{{\it TESS} TTVs}
\shortauthors{Hadden et al.}

\submitjournal{AAS Journals}

\begin{document}

\title{Prospects for TTV Detection and Dynamical Constraints with {\it TESS}}

\author[0000-0002-1032-0783]{Sam~Hadden}
\affiliation{Harvard-Smithsonian Center for Astrophysics, 60 Garden St., MS 51, Cambridge, MA 02138, USA}
\correspondingauthor{Sam~Hadden}
\email{samuel.hadden@cfa.harvard.edu}

\author[0000-0001-7139-2724]{Thomas~Barclay}
\affiliation{NASA Goddard Space Flight Center, 8800 Greenbelt Road,
Greenbelt, MD 20771, USA}
\affiliation{University of Maryland, Baltimore County, 1000 Hilltop Cir,
Baltimore, MD 21250, USA}

\author[0000-0001-5133-6303]{Matthew~J.~Payne} 
\affiliation{Harvard-Smithsonian Center for Astrophysics, 60 Garden St., MS 51, Cambridge, MA 02138, USA}

\author[0000-0002-1139-4880]{Matthew~J.~Holman}
\affiliation{Harvard-Smithsonian Center for Astrophysics, 60 Garden St., MS 51, Cambridge, MA 02138, USA}


\begin{abstract}
We consider the potential for the {\it Transiting Exoplanet Survey Satellite} ({\it TESS}) to detect transit timing variations (TTVs) during both its nominal and extended mission phases.
Building on previous estimates of the overall yield of planetary systems from the {\it TESS} mission, we predict that during its nominal two-year mission, {\it TESS} will observe measurable TTVs in {$\sim30$} systems, from which {$\mathcal{O}(10)$} planet will get precise mass measurements \emph{from TTVs alone}, {$\sim 5$} planets will have significant constraints placed on their masses from TTVs, 
and over a dozen systems will be singly transiting TTV systems.
We consider a number of different extended mission scenarios, and predict that in a typical scenario, an extended mission will allow {\it TESS} to increase the number of systems with measurable TTVs to a total of $\sim90$, from which {$\sim15$} planets will have precise mass measurements, {another $\sim 15$} will have significant constraints placed on their masses,
and $\sim60$ will be singly transiting TTV systems.
We also describe how follow-up transit observations of multiplanet systems discovered by the {\it TESS} mission can be optimally planned to maximize TTV mass and eccentricity constraints. 
\end{abstract}

\section{Introduction}
\label{SECN:INTRO}
    Transit timing variations (TTVs) are a powerful tool for measuring masses and eccentricities in planetary systems with multiple transiting planets \citep{Agol05,Holman05,Holman10}. The precise, long-term photometric observations of the \Kepler mission \citep{Borucki2010} allowed detection of timing variations in a large sample of transiting systems \citep{Mazeh2013,Holczer2016,Ofir18}. These timing variation measurements enable mass and eccentricity measurements of small planets in multitransiting systems \citep[e.g.,][]{JH2016,HL17}, often in cases that would otherwise be inaccessible to traditional radial velocity methods. NASA's {\it Transiting Exoplanet Survey Satellite} \citep[{\it TESS};][]{TESS15} represents the next generation in space-based photometric transit surveys and presents new opportunities for TTV mass and eccentricity measurements. Measuring masses for 50 planets smaller than $4~R_\earth$ is a level 1 science requirement of the {\it TESS} Mission.\footnote{\href{https://{\it TESS}.mit.edu}{https://{\it TESS}.mit.edu}} Therefore, it is of interest to know the degree to which TTV mass determinations can contribute mass measurements of {\it TESS} discoveries. 

    \citet{Ballard18} provides a simplified prediction for the expected TTVs among {\it TESS} discoveries by applying an empirically derived, multiplicity-dependent TTV probability from \citet{Xie2014}, estimating that $\sim$5\% of {\it TESS} discoveries will exhibit TTVs. However, \citet{Ballard18} does not consider the amplitude of such TTVs, whether they would be detectable with the timing precision derived from {\it TESS} photometry, and to what degree they can be used to derive dynamical constraints. The central aim of this work is to address these issues and estimate the expected yield of mass and eccentricity measurements from TTVs of {\it TESS} discoveries. 

   A significant contributory factor to the detection and utilization of TTVs in the \Kepler data was the observing strategy used by the mission: \Kepler stared continuously at the same relatively small ($\sim100$ sq. deg.) field of stars for $\sim4$ years, generating an approximately continuous data set of photometric observations for its $\sim 10^5$ target stars. In contrast, the {\it TESS} mission will cover the majority of the sky during its two-year mission, using 27 day-long stares at stripes of the sky covering $\sim 24^{\circ}\times96^{\circ}$. These stripes are then rotated about the ecliptic pole, generating a coverage pattern in which regions close to the ecliptic receive only 27 days coverage during the year, while only a small region near the ecliptic pole experiences continuous coverage. Full-frame image (FFI) data from each stripe are taken on a 30 minute cadence, while data for postage stamp regions of the specific pixels surrounding $\sim15,000$ targets per stripe are downloaded at two-minute cadence. 
   
    {\it TESS}'s significantly shorter time baselines, relative to {\it Kepler}, will make deriving dynamical constraints from TTVs more difficult for many multiplanet systems.    But short observational baselines do not necessarily preclude the possibility of extracting useful information from TTVs measured with {\it TESS}. For example, \citet{Rodriguez18} constrain two planetary masses using just nine TTV data points that cover a single planet-planet conjunction. Furthermore, \citet{Goldberg2018} show that it will be possible to combine {\it TESS}'s short time baseline observations with previous {\it Kepler} observations to refine dynamical constraints for a handful of TTV systems.

   An extension of the {\it TESS} mission beyond its nominal two years of operation could provide longer observational baselines and improve the opportunities for dynamical TTV constraints. \citet{Bouma17} and \citet{Huang2018arXiv} recently discussed the expected benefits of  various extended {\it TESS} mission scenarios in terms of the overall number of planet discoveries, finding that an extended mission could yield $\sim 2000$ new planet detections per additional year of operation. An important goal of this work is to quantify the expected improvements to TTV dynamical constraints under various extended mission scenarios.

    In this paper we investigate {\it TESS}'s expected yield of dynamical TTV constraints by generating a synthetic population of planetary systems, simulating their TTVs, and analyzing the dynamical information contained in their TTV signals. We explore how various extended mission scenarios improve the overall yield of dynamical constraints. We also investigate how follow-up transit observations of multitransiting systems can be effectively planned to optimize TTV dynamical constraints.

    This paper is organized as follows. In Section \ref{SECN:SYNTH} we describe the population of synthetic planets we use in our investigation, and the synthetic TTVs we ultimately derive from these. In Section \ref{SECN:TTV} we describe the TTV model we use to fit the synthetic TTV data. In Section \ref{SECN:RESULTS} we present the expected results from the nominal mission as well as various extended mission scenarios. 
    In Section \ref{SECN:FOLLOWUP}, we highlight the ability of our model to plan effective follow-up observations of transiting planets, even in systems that exhibit no appreciable TTVs during the nominal {\it TESS} mission.
    Finally in Section \ref{SECN:DISC} we discuss our results and conclusions.

\section{Synthetic Observations}
\label{SECN:SYNTH}
{\it TESS} began science operations in 2018 July. However, while the first few exoplanet discoveries are beginning to be announced \citep[e.g.][]{HuangPiMensae}, it will take several years until a reasonably large sample of planets and candidate planets is available. Therefore, for this work we need to synthesize a reasonable set of transit signals that might be expected to be detected by {\it TESS}.

\subsection{Synthetic Observations: Stellar and Planetary Population}
\label{s:synth:pop}

The generation of the synthetic population of planetary systems follows the same basic procedure employed in \citet{Barclay18}. We provide a broad outline of the procedure here and detail some modifications to the original procedure of \citet{Barclay18} that we make in order to more accurately model the architectures of multiplanet systems.  We refer the reader to the original text for more details. 

\citet{Barclay18} simulated planets around stars in version 6 of the {\it TESS} Input Catalog (TIC) Candidate Target List (CTL). There are a total of 3.8 million stars in this catalog and all have computed stellar properties such as mass, radius and brightness. Of these 3.8 million, 3.2 million were computed as observable to {\it TESS}, and were assumed to be observed by the mission using the 30-min integration time Full-Frame Image mode. From these, $214,000$ were assumed to be observed using the spacecrafts 2-min cadence mode. The stars observed at 2-min cadence were selected using the CTL prioritization metric \citep{Stassun2018}, but we forced the stars to be evenly distributed over the observing sectors. 

We modify the method of \citet{Barclay18} method for assigning planets to host stars in order to more accurately capture the architectures of multiplanet systems since planets' spacings strongly influence their TTVs. In particular, we aim to reproduce the period-ratio distribution of multiplanet systems inferred from {\it Kepler} using the debiased period-ratio distribution derived by \citet{Steffen15}.
As in \citet{Barclay18}, we begin by assigning each star in the sample a number of planets drawn from a Poisson distribution. $AFGK$ dwarfs were drawn assuming a mean of $\lambda=0.689$ for periods $\leq85$ days \citep{Fressin13}, while $M$ dwarfs are drawn assuming $\lambda=2.5$ for periods $\leq200$ days \citep{Dressing15}. Next, we assign periods and radii to the innermost planet from the measured period--radius distribution of planets from \citet{Fressin13} for $AFGK$ dwarfs and from \citet{Dressing15} for $M$ dwarfs. 
We differ from \citet{Barclay18} by drawing the periods of subsequent planets in the system from the period--ratio distribution in \citet{Steffen15}. Planets in a single system are drawn iteratively, the period ratio being with respect to the previous outermost planet at each step.
Radii of these subsequent planets are drawn from the appropriate period bins of the inferred period--radius distributions of \citet{Fressin13} or \citet{Dressing15}. Because giant planets almost never occur in multitransiting systems,  planets in multitransiting systems initially assigned radii larger than $6~R_\earth$ are reassigned new random radii from the period--radius distribution until a radius smaller than $6~R_\earth$ is drawn.

The eccentricities of the planets were assumed to follow a Rayleigh distribution with a scale parameter of $\sigma_e=0.02$, consistent with the eccentricity distribution of \Kepler multiplanet systems \citep{HL14,HL17,VanEylen15,Xie16}. The inclination of the inner-most planet was drawn from a distribution uniform in $\cos{i}$, where $i$ is the inclination with respect to the plane of the sky. All additional planets in any multiplanet system have {sky-projected inclinations drawn from a Gaussian distribution with $\sigma_i=2^{\circ}$ centered on the innermost planet's inclination.} This inclination selection strategy mimics the mutual inclination distribution seen in {\it Kepler}'s multis \citep{Fang12,Fabrycky14}.
{Because constraints on the intrinsic mutual inclination distribution of {\it Kepler} multis are uncertain we also simulate a second, low-inclination population with $\sigma_i=1.5^{\circ}$ to assess the influence of the assumed inclination distribution on our results. Results of our analysis for this low-$i$ population are described in Appendix \ref{SEC:APP:B}. All results in the main body of the paper pertain to our fiducial $\sigma_i=2^{\circ}$ population unless explicitly stated otherwise.} Finally, each planet is assigned a mass randomly drawn using the probabilistic mass--radius relationship code, {\sc FORECASTER}, of \citet{ChenKipping2017}.

We perform a basic check for stability of the multiplanet systems by determining whether each system's adjacent planet pairs satisfy the two-planet stability criterion of \citet{HL18}. We remove 10 systems from our sample that host adjacent planet pairs predicted to be chaotic and unstable. 

\subsection{Synthetic Observations: Detection Modeling and Transit Time Uncertainties}
\label{SECN:SYNTH:TRANS}

Having generated our sample of planetary systems, we use the detection model of \citet{Barclay18} to determine which stars host at least one planet that is expected to be detectable by {\it TESS}. In brief, the detection model of \citet{Barclay18} determines which planets will have transit detections with signal-to-noise ratio S/N$\ge 7.3$ based on their transit properties, the predicted {\it TESS} photometric noise level \citep[dependent on their host star's brightness; ][]{Stassun2018}, and the flux contamination reported in the CTL. Applying the detection model, we find a total of  {3756} stars host one or more planets with detectable transits. 
{Table \ref{tab:multi_tranet_summary} summarizes the multiplicity statistics of all systems in which at least one transiting planet is detected.}  Of these {3756 stars, 1814} host more than one planet. 
These multiplanet systems are the focus of our TTV study. 
We note that the stellar hosts of the multiplanet systems are split such that {$437$} are M dwarfs, and {$1377$} of earlier type. 

Of the {1814} multiplanet systems, {147} are multi-``tranet" systems \citep{Tremaine2012} containing two or more planets with detected transits while the other {1667} are single-``tranet" systems in which only one of the planets exhibit detectable transits while the other planets either do not transit or have transits below the S/N$\ge 7.3$ threshold. 
These systems are detected in both two-minute cadence data and thirty-minute cadence FFI data: of the {1814} multiplanet systems, {775} are in the two-minute cadence in the nominal mission, while {1039} are in the FFI data.

\begin{table}[]
    \centering
    \begin{tabular}{| c | c c c c | c|}
     & \multicolumn{4}{|c|}{$N$ transiting} & \\ 
 $N$ planets & 1 &	2 &	3 &	4& Total \\ \hline
1 &1942  & 0  & 0  & 0  & 1942  \\ 
2 &1024  & 64  & 0  & 0  & 1088  \\ 
3 &403  & 47  & 1  & 0  & 451  \\ 
4 &135  & 21  & 1  & 0  & 157  \\ 
5 &62  & 9  & 1  & 1  & 73  \\ 
6 &30  & 2  & 0  & 0  & 32  \\ 
7 &11  & 0  & 0  & 0  & 11  \\ 
8 &1  & 0  & 0  & 0  & 1  \\ 
9 &1  & 0  & 0  & 0  & 1  \\ 
\hline
Total &3609  & 143  & 3  & 1  & 3756 \\

    \hline
    \end{tabular}
    \caption{{Breakdown of the planet and transit multiplicities of systems with one or more detected transiting planet in the synthesized population.}}
    \label{tab:multi_tranet_summary}
\end{table}

In order to study what dynamical information may be gleaned from TTVs, we need an estimate of the precision with which {\it TESS} will be able to measure transit mid-times. 
Building on the work of \citet{Carter08,Carter09}, \citet{Price14} developed analytic expressions to approximate the uncertainties in various quantities derived from transit fitting as functions of planet properties and photometric precision. 
Their formula for the uncertainty in mid-transit times can be expressed as
\begin{equation}
  \sigma_{t_c} = \frac{\sigma}{\delta}\sqrt{\frac{\tau}{2}}\times \left\{
    \begin{array}{@{}ll@{}}
        \left(1-\frac{I}{3\tau}\right)^{-1/2}, & \text{if}\  \tau\geq I \\
        \left(1-\frac{\tau}{3I}\right)^{-1/2}, & \text{if}\  \tau  <  I 
    \end{array}\right.
    \label{eqn:uncert}
\end{equation} 
where 
$\sigma$ is the uncertainty of individual {\it TESS} photometric measurements,
$\delta$ is the planet's transit depth,
$I$ is the cadence (i.e., 2 minutes for pre-selected {\it TESS} targets and 30 minutes for planets discovered in the FFIs), and $\tau$ is the ingress time.
We use Equation \eqref{eqn:uncert} to estimate the expected transit mid-time uncertainty, $\sigma_{t_c}$, for each planet in our sample, by combining the photometric uncertainty, $\sigma$, of the host star with the transit depth, duration, and ingress times computed from the planet's radius and orbital properties. 
\citet{Goldberg2018} demonstrate that Equation \eqref{eqn:uncert} accurately predicts the precision of transit mid-time measurements when transit S/Ns are sufficiently high ($\text{S/N}\gtrsim 3$, see their Appendix A).

\subsection{Synthetic Observations: {\it TESS} Schedule / Coverage}
\label{SECN:SYNTH:COVERAGE}
A central aim of this work is to predict the expected improvements to the detection of TTVs in {\it TESS} data (and the mass measurements that can be extracted from these data) under various extended mission scenarios. We consider four distinct 3 yr extensions of the {\it TESS} mission, which we summarize in Table \ref{tab:Scen1}. Our extended mission scenarios cover two possible camera configurations:
the first, $C_4$, has camera 4 centered on the ecliptic pole as in the nominal mission. 
The second, $C_3$, has camera 3 centered on the ecliptic pole and provides a larger area of sky with multiple pointings, at the expense of coverage near the ecliptic equator. (\citet{Huang2018arXiv} refer to this configuration as `C3PO'.) 
We also consider two possible extended mission pointing sequences: one in which {\it TESS} remains pointed in the northern ecliptic hemisphere for the entire extended mission ($NNN$) and one in which {\it TESS} alternates hemispheres each year after starting in the north ($NSN$). 

We approximate the {\it TESS} sky footprint as a rectangular region spanning 
$24\deg\times 96 \deg $ in ecliptic longitude and latitude in order to determine which planets are observed during the various extended missions considered.

We ignore any potential TTVs of planets that are not detected during the nominal mission but later discovered during the extended mission.  We do not expect such planets to contribute significantly to {\it TESS}'s TTV yield for the same reasons they are missed during the nominal mission: either their transits are of low S/N and therefore their transit mid-times will be measured with poor precision or their orbital periods are longer than their nominal mission observational time baseline and they will have only a few transits during an extended mission.
\begin{deluxetable}{ll ll}
\tablecaption{Mission Scenarios 
\label{tab:Scen1}}
\startdata
\tablehead{
\multicolumn{2}{c}{Scenario} &
\multicolumn{2}{c}{Nominal} 
\\
\colhead{Symbol} &  
\colhead{Desc.} & 
\colhead{Pole Camera} &
\colhead{Pattern}
} 
$N$             &       Baseline  &  $C_4$ &  $SN$  \\
$E_{4,NSN}$     &       Extended &   $C_4$ &  $NSN$  \\
$E_{4,NNN}$     &       Extended &   $C_4$ &  $NNN$ \\
$E_{3,NSN}$     &       Extended &   $C_3$ &  $NSN$ \\
$E_{3,NNN}$     &       Extended &   $C_3$ &  $NNN$ \\
\enddata
\end{deluxetable}

\subsection{Synthetic Observations: {\it TESS} TTV Sample}
\label{SECN:SYNTH:TTVS}
{
To synthesize  TTVs, we use the analytic TTV model described below in Section \ref{SECN:TTV}.
We account for interactions between all planets with period ratios $P'/P<2.2$ in a given system, both transiting and nontransiting.
} 
Planets' initial mean anomalies are selected randomly from a uniform distribution. 
For simplicity, we assume all planets lie in the same plane in our dynamical integrations since the small mutual inclinations typical of multiplanet systems 
\citep[e.g.][]{Fang12,Figueira12,Fabrycky14}
have negligible influence on planets' TTVs \citep[e.g.,][]{HL16}. 
We integrate each simulation for 5 yr and produce the times of mid-transit for every transiting planet. 
In order to generate statistics we repeat 50 simulations for each system, drawing new random mean anomalies each time.

{
    We also compute a set of synthetic TTVs using the TTVFast $N$-body code of \citet{Deck14} which we analyze  in Appendix \ref{SEC:APP:B}. 
    These $N$-body-synthesized TTVs exhibit signs of strong resonant interactions much more frequently than is observed in the existing TTV planet population: mean motion resonances (MMRs) are rare among Kepler's population of multiplanet systems \citep{Fabrycky14} and very few Kepler TTV systems show the effects due to second- or higher-order resonances \citep{HL17}.
    The prevalence of these resonant effects in our synthesized TTVs likely owes to imperfect modeling the joint distribution of planetary systems' masses, period ratios, and eccentricities, upon which these  effects sensitively depend.
    While inferring a joint distribution of planet masses, period ratios, and eccentricities that successfully reproduces the prevalence of resonant effects in the observed population is a potentially fruitful avenue of future research, it is beyond the scope of this work.
    Instead, we opt to use the analytic model as a means of producing a TTV data set that is  similar to the observed TTV sample:
     our analytic model captures the effects of proximity to first order MMR and such TTVs comprise the majority of significant TTVs in the observed planet population \citep[e.g.,][]{Lithwick12}. 

     We expect our analysis  with the analytic model to provide an reliable estimate of how many planets are amenable to TTV mass measurements even though it may not faithfully reproduce every planet's TTV signal.
    This is because measurements of the chopping signal \citep{Nesvorny2014,DA15,HL16}, which our model accounts for, are usually necessary for deriving mass constraints from TTVs whether planets are near first-order MMRs,
    higher-order MMRs \citep[see, e.g., Kepler-26 b and c in ][]{HL16},
    or even librating in MMR \citep{Nesvorny2016}.
    Our analytic TTV model will faithfully reproduce the `chopping' component of the TTV as part of the $\delta t^{(0)}$ basis function component \citep[see ][]{Linial2018} even when the full TTV signal is not accurately modeled.}

\section{TTV Fitting}
\label{SECN:TTV}
    In this section we describe our methods for identifying significant TTVs in our simulated data and and extracting dynamical information. 
\subsection{TTVs as a linear combination of basis functions}
\label{SECN:TTV:Linear}
    Planets' TTVs can, to a good approximation, be treated as linear combinations of basis functions \citep{HL16,Linial2018}.  These basis functions depend only on planets' periods and are thus known when all dynamically relevant planets in a system transit. Specifically, the $n$th transit of the $i$th planet is given by 
\begin{multline}
    t_{i}(n)=nP_i + T_{i} + \\ \sum_{j\ne i} \mu_j\left( \delta t^{(0)}_{i,j}(n)+ ( {\cal Z}_{i,j}\delta t^{(1)}_{i,j}(n) + c.c. )\right)
    \label{eq:TTV:LINEAR:linear_model}
\end{multline}
    where $P_i$ is the $i$th planet's period, $T_i$ sets the time of the $i$th planet's first transit, $\mu_j$ is the planet--star mass ratio of the $j$th planet,
    \begin{equation}
        {\cal Z}_{i,j} \approx \frac{e_je^{i\varpi_j}-e_ie^{i\varpi_i}}{\sqrt{2}}
    \end{equation}
    is the combined complex eccentricity of planets $i$ and $j$ as defined in \citet{HL16} where $e_i$ and $\varpi_i$ are the eccentricity and argument of pericenter, `$c.c.$' denotes the complex conjugate of the preceding term, and $\delta t^{(0)}_{i,j}$ and $\delta t^{(1)}_{i,j}$  are TTV basis functions that capture the TTV induced by the $j$th planet. 
    It will frequently be convenient to refer to the real and imaginary components of $\delta t^{(1)}_{i,j}$ separately as $\delta t^{(1,x)}_{i,j}$ and $\delta t^{(1,y)}_{i,j}$, respectively. 
    A least-squares fit of a planet's TTV yields one real and one complex amplitude, $\mu_{j}$ and $\mu_{j}{\cal Z}_{i,j}$, for each perturbing planet.\footnote{Strictly speaking, the TTV model in Equation \eqref{eq:TTV:LINEAR:linear_model} is nonlinear because the TTV basis functions $\delta t^{(0)}$ and $\delta t^{(1)}$ depend on the planets' periods, $P$, and initial transit times, $T$. However, in most cases, planet periods and initial transit times can be determined iteratively, first determining periods by fitting planets' transit times with strictly linear ephemerides, then successively refitting transit times as combinations of linear ephemerides and the TTV basis functions constructed from planet periods determined at the previous iteration.}

    The TTV basis functions, $\delta t^{(0)}_{i,j}$, $t^{(1,x)}_{i,j}$ and $t^{(1,y)}_{i,j}$, can be computed using a perturbative solution of the planets motion \citep{HL16} or extracted from numerical integrations \citep{Linial2018}. Higher-order terms in ${\cal Z}_{i,j}$ can also be included as needed using the perturbative method described in \citet{HL16}. Details on computing the TTV basis functions are given in Appendix \ref{SECN:APP:A}.
    
A least-squares fit of a set of transit times provides the joint distribution of the basis-function amplitudes $\mu, \mu\ReZ,$ and $\mu\ImZ$ as a multivariate normal distribution. The quantities of interest in most applications will instead be  $\mu$ and $Z\equiv|\Z|$, who's joint distribution can be obtained from the integral
\begin{equation}
     P(\mu,Z) = \mu^2Z\int_{0}^{2\pi} N(\mu,\mu Z\cos\theta,\mu Z\sin\theta)d\theta\label{eq:prob_marginal}
 \end{equation}
 where $N(\cdot)$ is the Gaussian joint probability density of the TTV amplitudes, $\mu,\mu\ReZ,$ and $\mu\ImZ$, derived from the least-squares fit and the pre-factor of $\mu^2Z$ comes from the Jacobian of the variable transformation $(\mu,\mu\ReZ,\mu\ImZ)\rightarrow (\mu,Z,\cos\theta)$.
 
  TTV parameter inference is frequently hampered by a degeneracy between planet masses and eccentricities \citep{Lithwick12}. This is because, often, the basis functions $\delta t^{(0)}$ and $\delta t^{(1,x)}$ are both approximately sinusoids with identical phases. Consequently, there is a strong anticorrelation between the inferred amplitudes of these two basis functions, $\mu'$ and $\mu'\ReZ$.
   
   The TTVs of two or more planets are determined by fewer free parameters than assumed by a linear model approach. Consider the TTVs of an interacting pair of planets: the TTVs are determined by four free parameters (two planet--star mass ratios and two components of the planets' $\Z$). However, the linear model includes six independent TTV amplitudes: two planet-star mass ratios and two complex numbers of the form $\mu\Z$. Simply fitting two linear models to a pair of planets' TTV ignores the fact that the same combined eccentricity components appear in the basis-function amplitudes of both the inner and outer planet. Therefore, the linear model approach will likely overestimate the uncertainties in planet parameter constraints in some situations.
 \subsection{Estimating the precision of TTV planet parameter constraints}
\label{SECN:TTV:FOLLOWUP}
 The linear TTV model described above can be used to  predict the expected precision of planet parameter constraints that can be derived from a set of transit timing observations.  Let us  rewrite Equation \eqref{eq:TTV:LINEAR:linear_model} for the transit times of the $i$th planet in matrix form as
 \begin{eqnarray}
     \begin{pmatrix}t_i(1) \\ t_i(2) \\ \vdots \\ t_i(N_\text{trans.}) \end{pmatrix} = {\bf M}^{(i)}\cdot\begin{pmatrix}T_i \\ P_i \\ \mu_1 \\ \mu_1\ReZ_{i,1}\\ \mu_1\ImZ_{i,1} \\\vdots\end{pmatrix}
 \end{eqnarray}
 where ${\bf M}^{(i)}$ is a $[2+3N_\text{pert.}] \times N_\text{trans.}$ matrix for a planet with $N_\text{trans.}$ transit observations perturbed by $N_\text{pert.}$ companion planets.
 Let $\sigma_{t_c,l}$ be the uncertainty in the planet's $l$th transit time observation. We will assume that errors in transit time observations are normally distributed and independent. If we define the  design matrix as 
 \begin{eqnarray}
     A^{(i)}_{lm} = M^{(i)}_{lm}/\sigma_{t_c,l}~.
 \end{eqnarray}
 then the $[2+3N_\text{pl.}] \times [2+3N_\text{pl.}]$ covariance matrix between all basis-function amplitudes of the planet's TTV is given by 
 \begin{eqnarray}
    {\bf \Sigma} = [({\bf A}^{(i)})^T\cdot{\bf A}^{(i)}]^{-1}~.\label{eq:TTV:LINEAR:covariance}
 \end{eqnarray}
 Importantly, ${\bf \Sigma}$ does not depend on the values of the basis-function amplitudes. Only knowledge of the planets' periods and initial orbital phases, which come directly from observations, is required to determine the TTV basis functions, and hence ${\bf \Sigma}$. 
 Therefore, it is possible to predict the expected precision of TTV amplitude measurements for a set of transit observations without knowledge of the masses and eccentricities of the planets in the system.
 
\begin{figure}
    \centering
    \includegraphics[width=\columnwidth]{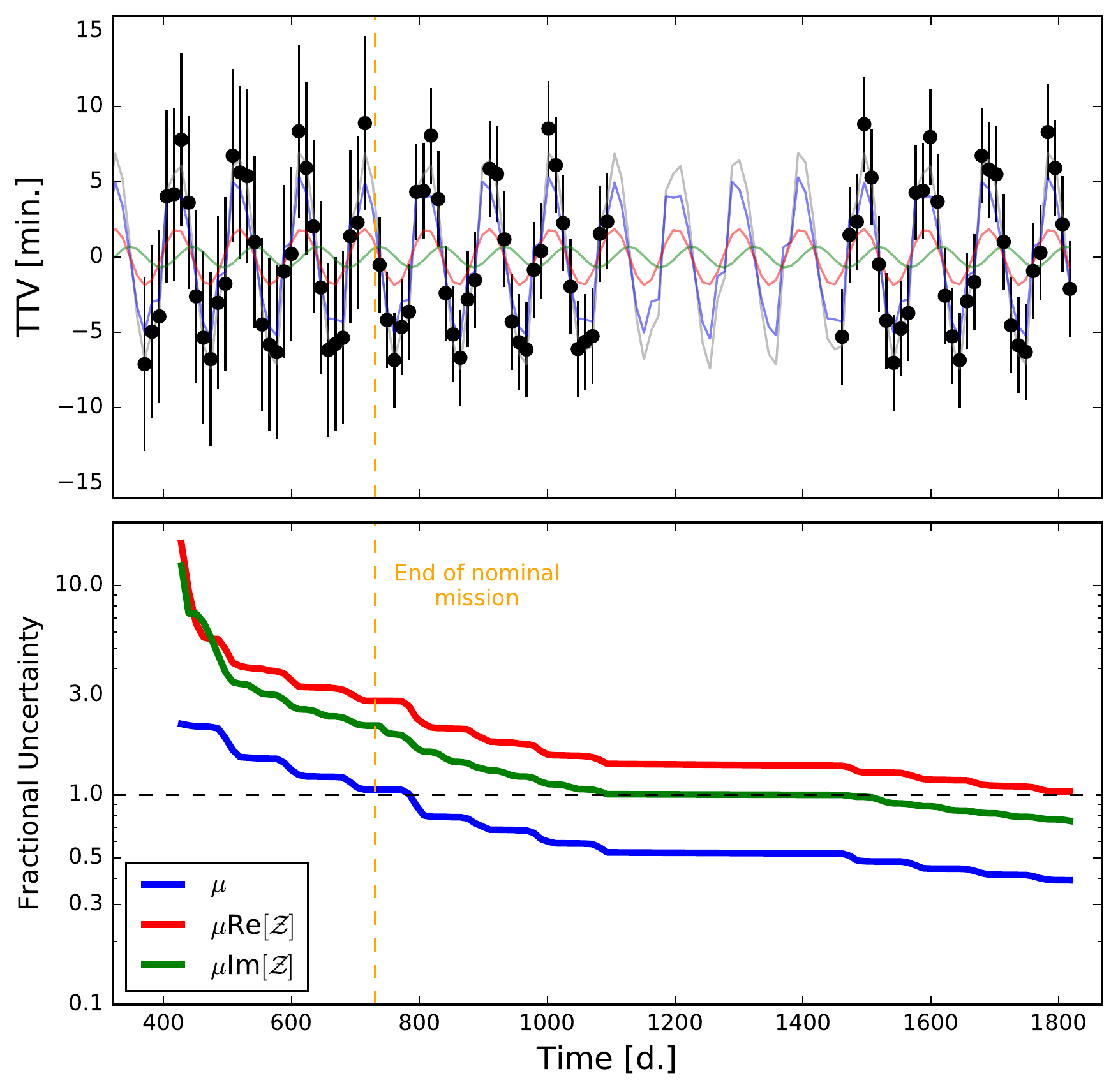}
    \caption{
    {Top panel:} simulated TTV of a planet on a 11.5 day orbit near a 2:1 MMR with an exterior 37$M_\oplus$ perturber. Black points show the results of an $N$-body integration using the TTVFast code \citep{Deck14}. The analytic TTV model, Equation \eqref{eq:TTV:LINEAR:linear_model}, is shown in gray with its constituent basis functions plotted as colored curves: $\delta t^{(0)}$ in blue, $\delta t^{(1,x)}$ in red, and $\delta t^{(1,y)}$ in green.
    {Bottom panel:} decrease in the fractional $1\sigma$ uncertainties of the the TTV model basis-function amplitudes with an increasing number of observed transits.  The black dashed line marks a fractional amplitude uncertainty of unity.
    \label{fig:TTV:LINEAR:projected_uncertainties}
    }
\end{figure}
    Figure \ref{fig:TTV:LINEAR:projected_uncertainties} illustrates the application of the linear TTV model with an example TTV data set taken from our synthesized {\it TESS} sample. 
    The top panel shows the TTV of a planet near a 2:1 MMR with an exterior $37 M_\oplus$ perturber, along with the decomposition of the TTV into its constituent basis functions. This particular system is located in the continuous viewing zone of the northern ecliptic hemisphere and is observed at a thirty-minute cadence during the nominal mission and then receives two-minute cadence observations during the first and third year of an extended mission. In the bottom panel we plot the predicted fractional uncertainty in the basis-function amplitudes versus the observing baseline. The amplitude uncertainties are determined by computing appropriate diagonal elements of $\Sigma$ in Equation \eqref{eq:TTV:LINEAR:covariance}. In this example, the additional observations provided by the extended mission lead to a measurement of the perturbing planets' mass with a fractional uncertainty of $\sim 40\%$ by constraining the amplitude of the $\delta t^{(0)}$ basis function plotted in blue in the top panel of Figure \ref{fig:TTV:LINEAR:projected_uncertainties}.

\section{Results}
\label{SECN:RESULTS}
We search the synthetic catalog of transit times (described in Section \ref{SECN:SYNTH}) for systems that show significant TTVs. 
We begin by reducing our synthetic transit time catalog to those systems that could exhibit detectable TTVs under any of the mission scenarios considered. To do so, we reduce the catalog to those systems in which one or more planets shows significant timing residuals, after removal of the best-fit linear ephemeris from the full 5-year simulated data set.\footnote{In reality, no star is observed for the entire 5 yr data set under any of the mission scenarios, so this is a conservative selection criterion.}
After computing the $\chi^2$ values of timing residuals, we retain
systems containing at least one planet with timing residuals that are inconsistent with Gaussian noise at $>95\%$ confidence in multiplanet systems or $>99\%$ in singly transiting systems. (A lower threshold is applied to multiplanet systems since dynamically-induced variations must exist at some level due to the the presence of confirmed companions.)
When computing $\chi^2$ values, timing uncertainties for years 3--5 are assigned assuming 2 minute cadence observations. under the assumption that anything discovered in the nominal missions will subsequently be transitioned to 2 minute cadence.

\subsection{Multitranet systems}
\label{SECN:RESULTS:multi}

\begin{figure*}[]
    \centering
    \includegraphics[width=0.95\columnwidth]{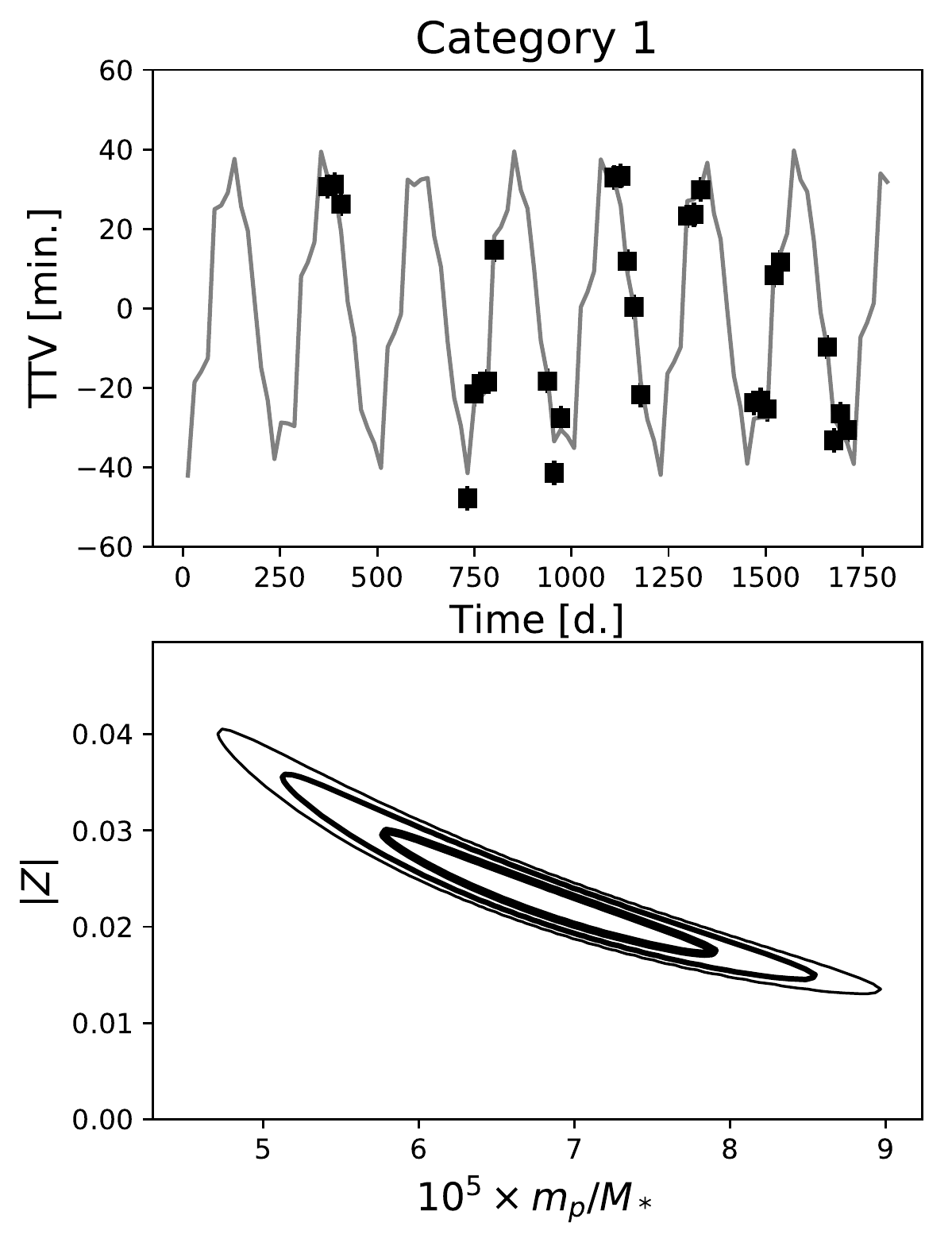}
    \includegraphics[width=0.95\columnwidth]{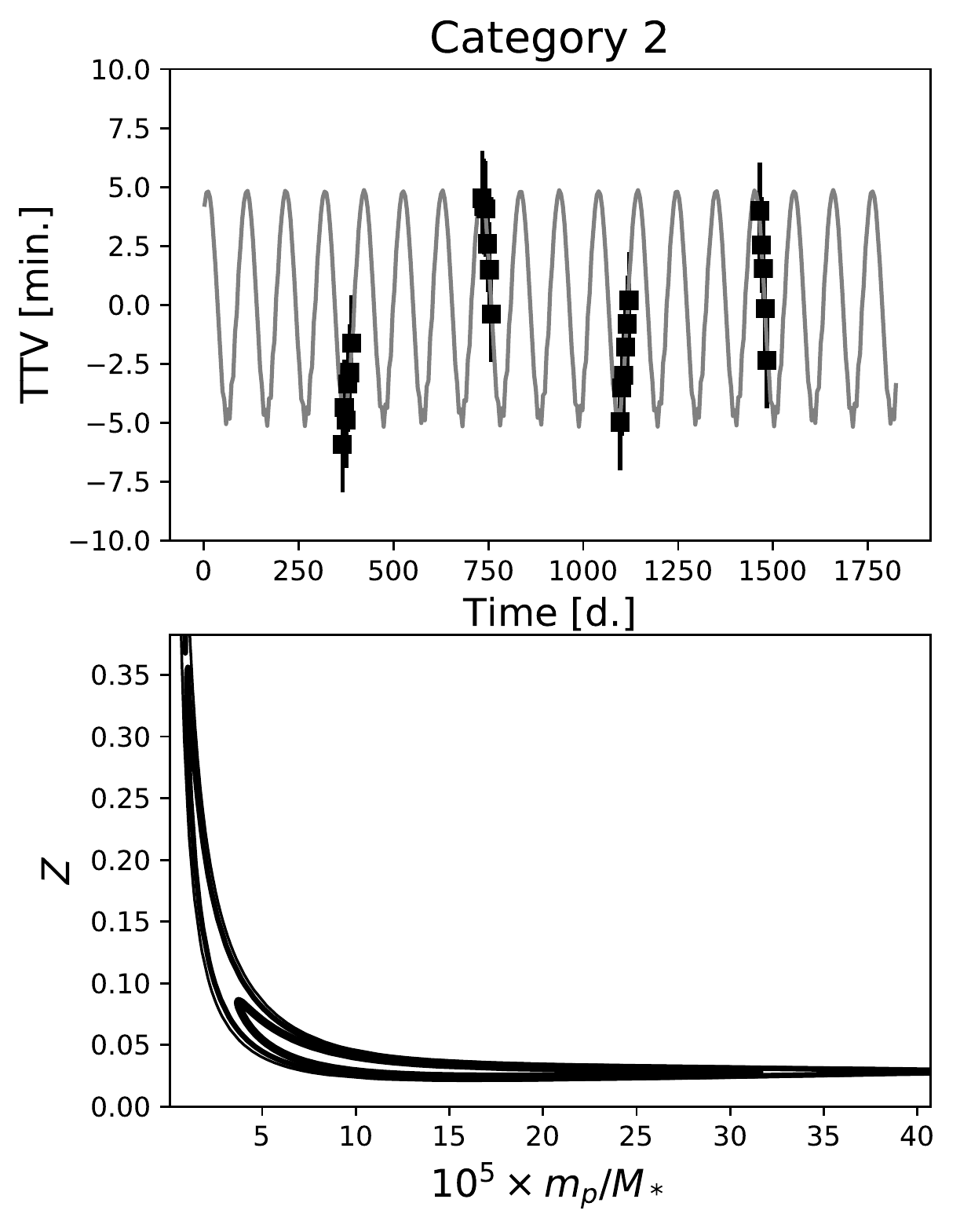}
    \caption{Two example TTVs representative of Category-1 (left) and Category-2 (right) as defined in Section \ref{SECN:RESULTS:multi}. The top panels show TTVs plotted versus time in black with the $1\sigma$ transit mid-time uncertainties indicated by error bars. The best-fit analytic TTV models are also plotted in gray. The bottom panels show joint constraints derived from the TTVs for the planet-star mass ratio and the combined eccentricity, $|{\cal Z}|$, by plotting  $1,2$ and $3\sigma$ contours derived using Equation \eqref{eq:prob_marginal}. Unambiguous detection of the $\delta t^{(0)}$ TTV amplitude in the left-hand panel leads to a strong constraint on the perturbing planet's mass while a strong degeneracy between  $\delta t^{(0)}$ and $\delta t^{(1,x)}$ amplitudes translates to a strong degeneracy between perturber mass and the planet pair's combined eccentricity in the right-hand panel.}
    \label{fig:system_categories_example}
\end{figure*}

After our initial selection of possible TTVs (described above), we attempt to identify multi-tranet systems with significant TTV signals, in which the TTVs can be attributed to interactions among the observed planets. 
Such systems are generally the richest targets for TTV analysis since mass measurements can yield density constraints for the observed planets.  To identify these systems, we fit the transit times of every multi-tranet TTV system in our catalog via least-squares with the linear model described in Section \ref{SECN:TTV}.  Interactions between planets separated by a period ratio greater than $P'/P>2.2$ are ignored in the least-squares fits. These linear models account only for those planets that are observed to transit and ignore any potential nontransiting or undetected perturbers.
We do not fit the TTVs of planets for which our analytic TTV model is underdetermined due to insufficient transits.%
\footnote{The analytic model is underdetermined for planets with fewer than $2+3N_\text{pert}$ transits. We still account for the perturbations of such planets on their companions showing a sufficient number of transits.} After fitting a multi-tranet system's TTVs with our linear model, it falls into one of three categories:
\begin{enumerate}
    \item {\bf Category 1} systems yield direct mass measurements. The system hosts at least one planet for which a $\delta t^{(0)}$ TTV component amplitude is non-zero at $>1\sigma$ confidence. This means that one or more planet's TTVs constrain the mass of its perturbing companion(s). 
    \item {\bf Category 2} systems exhibit significant TTVs, but  mass inference is hampered by a mass-eccentricity degeneracy.  For one or more planets in the system, the error ellipse defined by best-fit TTV amplitudes and their co-variance matrix excludes 0 at $>2\sigma$ confidence. However, all $\delta t^{(0)}$ amplitudes are consistent with 0, so the planets' masses are poorly constrained and strongly degenerate with the eccentricities. 
    \item {\bf Category 3} systems show significant timing residuals but are poorly fit by the analytic model. The planets timing residuals, after subtracting off the best-fit analytic model, indicate a poor fit which we define as a $\chi^2$ value inconsistent with random Gaussian noise at $>3\sigma$ significance.
    {For TTVs synthesized with the analytic model, poor fits are caused either by the presence of unseen additional perturbers or where the analytic model is under-determined due to an insufficient number of transits. For the TTVs synthesized via $N$-body simulation analyzed in Appendix \ref{SEC:APP:B}, this category also includes TTVs exhibiting effects not captured by the analytic TTV model (such as proximity to second- or higher-order resonances).} In some instances of this latter case, it may be possible to extract dynamical mass constraints from the TTVs. 
\end{enumerate}

{
Our categories are mutually exclusive, meaning that systems satisfying the criteria of multiple categories simultaneously are included in only one of them. Any system satisfying the Category 3 criterion is automatically classified as Category 3 irrespective of whether it also satisfies the criteria of either of the other two categories.
In principle, a TTV system may or may not simultaneously meet the criteria for classification in Categories 1 and 2: a TTV can have $\delta t^{(0)}$ amplitude detected with a $ >1\sigma$ significance (Category 1) independent of whether the total TTV signal is non-zero at  $>2\sigma$ confidence (Category 2).  We find that, in practice, nearly every Category 1 TTV also satisfies the criterion for inclusion in Category 2 and we classify such systems as Category 1. Additionally, all Category 1 and 2 systems exhibit $>3\sigma$ timing residuals relative to a linear ephemeris, and would therefore be included in Category 3 if their interactions were not modeled by the analytic model (though this is not an explicit requirement of our categorization scheme).
}

\begin{figure*}[htp]
    \begin{minipage}[b]{\textwidth}
        \centering
        \includegraphics[width=\textwidth]{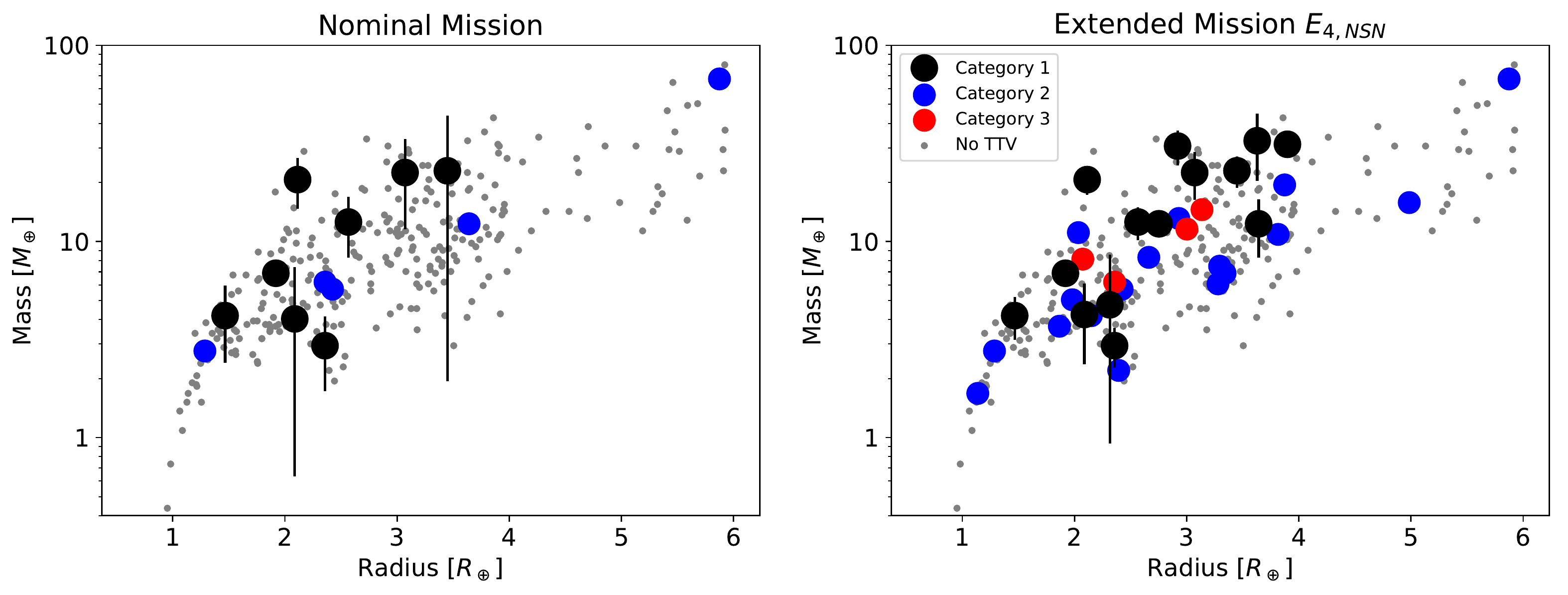}
        \caption{
        Mass--radius diagram of planets in our simulated sample of multitransiting systems. 
        Large points with $1\sigma$ error bars show measurements recovered from TTV fitting.
        The different colors indicate the categories defined in Section \ref{SECN:RESULTS:multi}.
        Planets with a well measured mass derived from TTV in Category 1 are plotted in black.
        Planets that induce a significant Category 2 TTV in a companion planet(s), but do not have recoverable masses, are indicated in blue. Companions to planets with Category 3 TTVs are plotted in red. 
        For Category 3 TTVs, only transiting companions with period ratios of $P'/P<2.2$ are plotted in red in order to exclude cases where the significant residuals are caused by intervening undetected perturbers. All other detected planets that do not induce measurable TTVs are plotted in gray.}
        \label{fig:massradius}
    \end{minipage}
\end{figure*}

Figure \ref{fig:system_categories_example} shows representative TTVs from the first two categories. For planets in Category 1, the TTV will provide a measurement of the perturbing companion's mass as well as the planet pair's combined eccentricity.
Planets in Category 2 exhibit a degeneracy between the perturber mass and eccentricity. An independent measurement of the perturber's mass, e.g., via radial velocity, can yield eccentricity constraints in such systems. 

Figure \ref{fig:massradius} plots the masses and radii of the TTV detections from one of the 50 realizations of our synthetic TTV catalog. The left-hand panel shows the results of the nominal mission: 
{eight} mass measurements derived from Category 1 TTVs as well as {five} Category 2 TTVs that provide a degenerate joint mass-eccentricity constraint similar to the example in the right-hand panel of Figure \ref{fig:system_categories_example}. The masses and radii of all additional detected planets in multitransiting systems are plotted in gray. 

The right-hand panel of Figure \ref{fig:massradius} shows the mass-radius results for the same realization of the TTV catalog after three additional years of an $E_{4,NSN}$ extended mission. The extended mission yields {six} additional Category-1 mass measurements and improves the precision of the mass measurements obtained during the nominal mission. The extended mission also adds {12} new Category-2 planets along with {four} Category-3 systems. 
All but two of the detected TTVs are induced by planets smaller than $4R_\earth$.

Table \ref{tab:ttv_summary} summarizes the predicted yields of detected TTV systems, separated into the above classifications, in both the nominal and extended mission scenarios.\footnote{
{
    A pre-print version of this paper posted to \href{https://arxiv.org/abs/1811.01970v1}{arXiv.org} in 2018 November projected somewhat different TESS TTV yields. 
    Most significantly, the number of planet mass measurements expected from the nominal mission has increased from an original prediction of only $\sim 1$ to nearly  $\sim 10$. 
    There are two principal reasons for the increased number of predicted mass measurements: 
    first, we corrected our method for generating mutual inclinations. Previously, sky-projected inclinations were drawn from a Rayleigh distribution with $\sigma_i=2^\circ$ whereas now they are drawn from a Gaussian distribution. This results in slightly flatter systems and has increased the number of multitranet systems from 136 to 147 (i.e., 11 more systems) in our fiducial $\sigma_i=2^\circ$ population.
    Second, and more significant, we now synthesize our TTV population by means of the analytic model, rather than $N$-body simulation. 
    As discussed in Section \ref{SECN:SYNTH:TTVS}, $N$-body simulations of our synthetic population tended to produce an excessive number of TTVs showing indications of strong resonant interactions. Many such systems' TTVs, when simulated instead with the analytic model, become Category 1 systems, thus increasing the predicted yield of mass measurements.
    Comparing Table \ref{tab:ttv_summary} with Table \ref{tab:ttv_summary_Nbody} in Appendix \ref{SEC:APP:B:Nbody} serves to illustrate the significance of this effect. 
    Notably, the range of expected mass measurement forecast in Table \ref{tab:ttv_summary_Nbody} (2--9 during a nominal mission) is similar to the 0 to 7 mass measurements projected in the original arXiv version.
    Additionally, we have corrected some errors in our analysis pipeline that caused the numbers of Category 3 systems and significant TTVs of singly transiting planets to be miscounted.
}} We list the median, minimum, and maximum number of planets falling within each category from the 50 simulation iterations randomized in planets' initial orbital phases.
Overall, the various extended mission scenarios yield similar numbers of TTV detections.

\begin{table*}[]
    \centering
    \begin{tabular}{ c | c c c c |}
    	& Category 1: & Category 2:  & Category 3: & Singles \\  
	& Mass Measured & Significant, No Mass & Residuals &  \\  
 \emph{Mission}  & Med. (min., max.) & Med. (min., max.) & Med. (min., max.) & Med. (min., max.) \\ \hline 
Primary & 9 (6, 11) & 5 (4, 8) & 2 (1, 4) & 15 (10, 19)\\
$E_{3,NNN}$ & 10 (9, 12) & 12 (9, 14) & 5 (3, 7) & 41 (36, 45)\\
$E_{3,NSN}$ & 11 (9, 13) & 14 (11, 16) & 7 (5, 8) & 53 (48, 59)\\
$E_{4,NNN}$ & 14 (13, 16) & 13 (11, 15) & 7 (5, 9) & 53 (46, 58)\\
$E_{4,NSN}$ & 16 (13, 18) & 17 (13, 20) & 8 (6, 9) & 63 (56, 67)\\

    \hline
    \end{tabular}
    \caption{Summary of the expected TTV yields for different {\it TESS} mission scenarios as defined in Section \ref{SECN:SYNTH:COVERAGE}. Listed are the median, minimum, and maximum number of planets,taken from the  50 iterations of our synthetic TTV catalog, that fall within each of categories defined in Section \ref{SECN:RESULTS:multi} as well as the number of singly transiting planets exhibiting significant TTVs.}
    \label{tab:ttv_summary}
\end{table*}
%

\subsection{Single-tranet TTVs}
\label{SECN:RESULTS:single}
    In addition to the multitransiting TTV systems described above, {\it TESS} is expected to discover a number of singly transiting planets that exhibit TTVs. While such TTVs generally offer little information beyond an indication of additional planets in the system, it may be possible to extract dynamical constraints in some instances \citep[e.g.,][]{KOI142}. 
    Table \ref{tab:ttv_summary} summarized the number of such systems whose transit timing residuals are inconsistent with Gaussian random noise at $>3\sigma$ for  various mission scenarios. In all {extended} mission scenarios, the expected number of singly transiting TTV planets outnumbers the total combined multitranet systems in Categories 1--3. 
    
\begin{figure}
    \centering
    \includegraphics[width=\columnwidth]{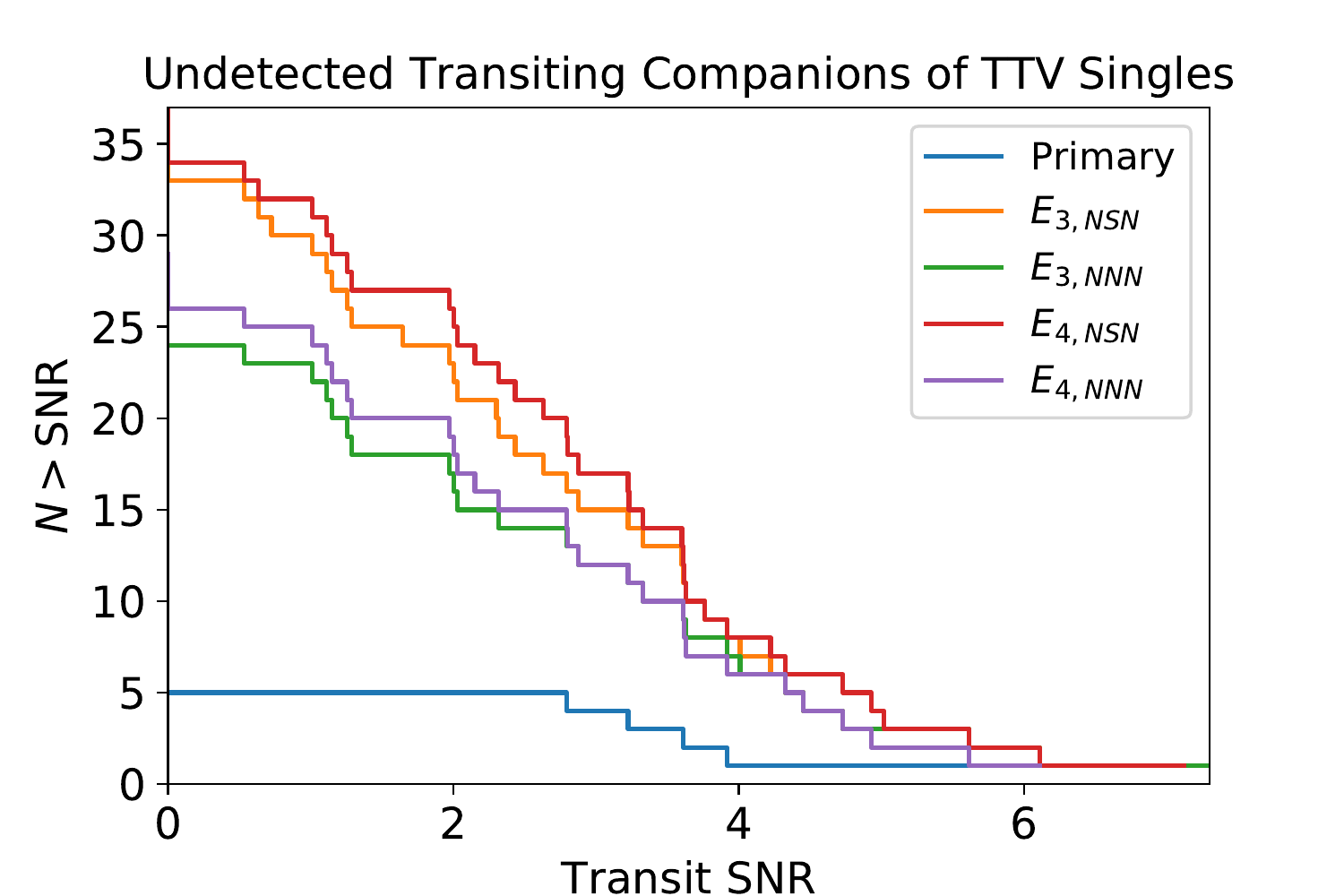}
    \caption{The transit SNR of transiting companions to single-tranet TTV planets. Different color curves correspond to the different mission scenarios described in Section \ref{SECN:SYNTH:COVERAGE}. Additional transits that occur during the extended mission are \emph{not} included in the plotted transit SNR.}
    \label{fig:single_companions}
\end{figure}

    Many of the multiplanet systems hosting single-tranet TTV planets contain additional transiting planets whose transits fall below {\it TESS}'s $7.3$ S/N detection criterion \cite[e.g.][]{Sullivan2015}. Figure \ref{fig:single_companions} shows the transit S/Ns of undetected transiting companions to the single-tranet TTV planets discovered in various mission scenarios. Many such companions could be recoverable upon relaxing the $7.3$ transit S/N requirement. If these planets are responsible for the TTV of the initially detected planet then the TTVs could yield dynamical constraints on the mass and eccentricities of these planets.

\section{Planning follow-up transit observations}
\label{SECN:FOLLOWUP}
In section \ref{SECN:RESULTS} we summarized the yield of TTV measurements based solely on transits times derived from {\it TESS} photometry. Here we briefly consider how improved dynamical constraints can be obtained by combining {\it TESS} data with follow-up transit observations from other space- or ground-based facilities. 
The linear TTV model's ability to project the expected precision of planet parameter constraints is particularly useful in planning effective follow-up observations of multiplanet systems. 
We demonstrate that follow-up observations can be planned to successfully extract mass measurements even for systems that exhibit no appreciable TTVs during the nominal {\it TESS} mission. This is best illustrated by way of an example. 

\begin{figure}
    \centering
    \includegraphics[width=0.95\columnwidth]{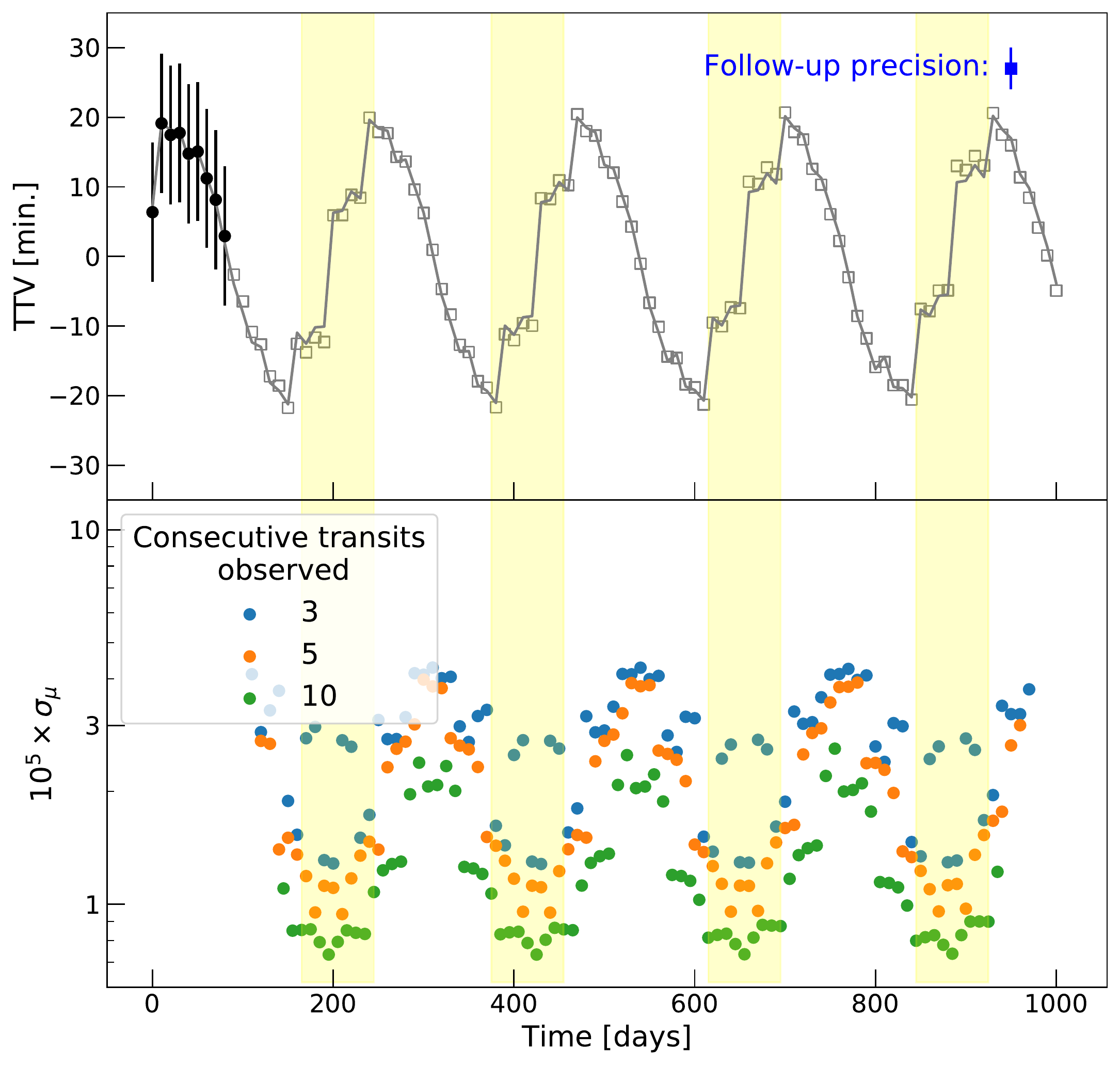}
    \caption{{Top panel:} TTV of a planet on a 10 day orbit perturbed by an external companion with planet--star mass ratio of $\mu=3\times10^{-5}$ near a 4:3 MMR. The black points show transits observed during three sectors of {\it TESS}'s nominal mission with error bars indicating a ten-minute transit mid-time uncertainty. Gray squares show additional transits after the {\it TESS} mission observations and the gray line shows the best-fit analytic TTV model. {Bottom panel:} the precision in the determination of the perturbing planet's mass achieved with various sets of follow-up transit observations. Specifically, each point shows the standard deviation in the  perturbing planet's planet-star mass ratio, $\sigma_\mu$ when $N=3,5,$ or 10 consecutive transits, centered on the plotted time, are observed with three-minute transit mid-time precision.  Optimal times for transit observations are highlighted in yellow.}
    \label{fig:follow_example}
\end{figure}

The top panel of Figure \ref{fig:follow_example} shows the TTV of a hypothetical planet on a 10 day orbit perturbed by an exterior perturber near a 4:3 MMR with a mass of $3\times 10^{-5}M_*$ and a combined eccentricity of $Z\approx0.01$. The system is observed for three successive {\it TESS} sectors (82.2 days) during the nominal mission. The  transits observed during the nominal mission are sufficient to precisely determine the periods and initial transit times of the planet pair but do not yield dynamical constraints on planet masses and eccentricities. 
In fact, during the nominal {\it TESS} mission, these planets have such small timing residuals that they do not meet our criteria for selection in any of the categories described in Section \ref{SECN:RESULTS}.
However, the  periods and initial transit times determined from the {\it TESS} observations fully determine the TTV basis functions, $\delta t^{(0)}$ and $\delta t^{(1)}$. The value of these basis functions at any proposed set of future follow-up transit observations, along with the expected mid-time precision of the follow-up observations, are sufficient to compute a covariance matrix, $\Sigma$, defined in Equation \ref{eq:TTV:LINEAR:covariance}. This covariance matrix determines the precision with which the TTV basis-function amplitudes are expected to be measured, and thus, the expected precision with which the perturbing planet's mass and the combined eccentricity would be constrained.

The bottom panel of Figure \ref{fig:follow_example} shows  predictions for how precisely the perturbing planet's mass can be determined if additional transit observations are obtained. Each point in the bottom panel shows the expected standard deviation in a measurement of the $\delta t^{(0)}$ basis-function amplitude, corresponding to the perturbing planet's planet--star mass ratio, for hypothetical sets of follow-up observations of consecutive transit taken at different times. The different color points correspond to follow-up observations spanning 3, 5, or 10 consecutive transits measured with three-minute mid-time uncertainty.  Optimal times for follow-up, i.e., those that yield the smallest resultant $\sigma_\mu$ values, are highlighted. These optimal times clearly correspond to locations in the TTV signal where a short-period chopping signal is evident. We emphasize that these optimal follow-up times can be predicted from the nominal mission data alone.

This example serves to illustrate the potential importance of photometric follow-up observations to the overall TTV yield of {\it TESS}. In this particular example, a follow-up observation spanning just five judiciously chosen transits of the planet would be sufficient to yield a measurement of the perturber's mass at a significance of $\sim 3\sigma$ despite a lack of significant TTV signal over the course of the nominal {\it TESS} mission. 
\section{Discussion}
\label{SECN:DISC}
 The main goal of this work has been to provide a quantitative estimate of {\it TESS}'s potential to detect transit timing variations (TTVs) and determine the dynamical constraints on planet masses and eccentricities that can be derived from these variations. We summarize our findings as follows.
\begin{enumerate}
\item{
The {\it TESS} mission's TTV yield after completion of its nominal mission will likely be modest, providing only a handful of significant TTV signals in multitransiting systems, of which {$\mathcal{O}(10)$} can be expected to yield a direct measurement of a planet's mass, not subject to a mass-eccentricity degeneracy (i.e., a Category-1 TTV, as defined in Section \ref{SECN:RESULTS:multi}). Additionally, we expect approximately a dozen singly transiting planets with significant TTVs will be identified in the nominal mission.
}
\item{ 
A three-year extension of the {\it TESS} mission can significantly enhance the number of TTV systems that provide dynamical constraints. We expect an extended mission to yield $\sim 30$ significant TTVs in multitransiting systems (i.e., the total number in Categories 1, 2, and 3 in Table \ref{tab:ttv_summary}), of which {$\gtrsim 10$} will yield a direct planet mass measurement. The number of significant TTVs is relatively insensitive to the choice of extended mission observing strategy. In addition, {$\sim40-60$} TTVs of singly transiting planets will be identified. Such TTVs will, in a number of instances, reveal transiting companions that do not meet the formal 7.3 S/N threshold for detection but that nonetheless may be recovered from the light curve (Figure \ref{fig:single_companions}).
}
\end{enumerate}
One of the primary science objectives of the {\it TESS} mission is to determine masses for 50 planets smaller than $4R_\earth$. An extended mission could allow TTV mass measurements to contribute significantly toward this goal.

In addition to projecting the yield of dynamical constraints derived from {\it TESS} observations alone, we described how follow-up transit observations of discoveries can be optimally planned to yield dynamical information. 
The ultimate contribution of such observations to the overall TTV yield of {\it TESS} will depend on the availability of follow-up resources, both ground- and space-based, and the photometric precision that can be achieved.

TTV signals have a complicated and sensitive dependence on planetary systems'  orbital architectures. Therefore, imperfect knowledge of the underlying distribution of multiplanet system architectures is a source of uncertainty in our predicted yields.  We have attempted to use the best available constraints from the literature when drawing planet properties such as radii, masses, periods, and eccentricities to generate our synthetic planetary system population. However, possible correlations between these parameters will influence our expectations for the {\it TESS} mission's TTVs and remain poorly quantified. Recent works based on the sample of \Kepler multiplanet systems have shown that planets in the same system tend to be similar in size \citep{Weiss2018} as well as mass \citep{Millholland2017},  and \citet{Zhu2018} find evidence that lower multiplicity systems tend to be dynamically hotter. Our synthetic planet population has not attempted to incorporate these correlations.  Our lack of knowledge is especially acute for  systems orbiting around M dwarfs, which are underrepresented in the \Kepler sample \citep[e.g.][]{Huber16}.

The methodology we develop in this work can be applied to plan follow-up transit observations to constrain planet masses as well as to evaluate the TTV yield of observing strategies of future transit survey missions such as PLATO \citep{PLATO}. The linear TTV model should also prove useful for exploring joint constraints attainable from the combination of TTV and RV measurements. For example, independent mass constraints from RV measurements can break the mass-eccentricity degeneracy in a TTV system so that the TTVs yield a direct measurement of combined eccentricity, $Z$ \citep[e.g.,][]{Petigura2018}.  {As a rough estimate of how many systems may be amenable to joint characterization by TTV and radial velocity, we find only 2 multi-transiting systems in our synthetic population around host that: (a) show significant TTVs during the nominal mission, (b) host at least one transiting planet with a radial velocity semi-amplitude greater than a meter per second, and (c) orbit stars with {\it TESS} magnitudes brighter than 11. When the various extended mission scenarios are considered, this increases to between 6 to 12 systems.
}
Python code for computing the linear TTV model basis functions and fitting transit time observations is available at \href{https://github.com/shadden/TTV2Fast2Furious}{github.com/shadden/TTV2Fast2Furious}.
{Our synthesized planet populations are available online at \href{https://doi.org/10.5281/zenodo.2852173}{doi.org/10.5281/zenodo.2852173}}.


\section*{Acknowledgments}
{
We thank Darin Ragozzine for his thorough and insightful referee report. 
This project was  developed in part at the Building Early Science with {\it TESS} meeting, which took place in 2019 March at the University of Chicago.}
This work has made use of the TIC, through the {\it TESS} Science Office's target selection working group (architects K. Stassun, J. Pepper, N. De Lee, M. Paegert, R. Oelkers).
MJH and MJP gratefully acknowledge 
NASA grants NNX12AE89G, NNX16AD69G, and NNX17AG87G, as well as support from the Smithsonian 2015-2017 Scholarly Studies program.
SH gratefully acknowledges the CfA Fellowship. TB was supported by the Sellers Exoplanet Environments Collaboration at Goddard Space Flight Center.
The computations in this paper were run on the Odyssey cluster supported by the FAS Science Division Research Computing Group at Harvard University.
\software{TTVFast \citep{Deck14}, FORECASTER \citep{ChenKipping2017}, Matplotlib \citep{Hunter2007}, NumPy \citep{numpy}, Pandas \citep{pandas}, SciPy \citep{scipy}}


\appendix
\section{Alternative Synthetic Planet Populations}
\label{SEC:APP:B}
{
In this appendix we list TTV results for a synthetic low-$i$ planet population generated with a mutual inclination dispersion of $\sigma_i=1.5^{\circ}$ rather than $\sigma_i=2^{\circ}$ used in the body of the paper. We also compare our fiducial TTV catalog synthesized with the analytic TTV formulas described in Section \ref{SECN:TTV} to a catalog synthesized via $N$-body integration. 

\subsection{Low-$i$ population}
\label{SEC:APP:B:LowI}
Table \ref{tab:multi_tranet_summary_low_i} summarizes the planet and tranet multiplicities of systems in the low-$i$ population. As expected, there are more multitransiting systems in the low-$i$ population, with 23 ($=166-143$) more two-tranet systems and 4 ($=7-3$) more three-tranet systems. Table \ref{tab:ttv_summary_low_i} summarizes TTV detections and dynamical constraints for the low-$i$ population. The median number of multitransiting TTV detections of each type increases slightly for the low-$i$ (with a corresponding decrease in the number of singly transiting TTV) though the ranges produced by randomizing over orbital phases largely overlap between our fiducial and low-$i$ populations.

\subsection{TTVs Synthesized with $N$-body}
\label{SEC:APP:B:Nbody}
    In Section \ref{SECN:SYNTH:TTVS}  we synthesized planets' transit timing variations using analytic formulas. 
    We also have run simulations that instead use the TTVFast code of \citet{Deck14} to generate transit times via $N$-body integration. 
    Table \ref{tab:ttv_summary_Nbody} summarizes TTV detections and dynamical constraints derived from TTVs synthesized via $N$-body simulation. 
    As shown by Table \ref{tab:ttv_summary_Nbody}, the $N$-body sample contains a significant number of Category 3 systems poorly fit by the analytic model. 
    All Category 3 TTVs generated with the analytic formulas are poorly fit due to the presence undetected perturbers inducing TTVs that are unaccounted for when fitting them with the analytic model.
    By contrast, $N$-body TTVs with significant residuals can be caused by undetected perturbers {as well as} failures of the analytic formulas to accurately model the planets' interactions.
    The analytic model fails to accurately capture TTVs of planets that are strongly affected by a second- or higher-order MMRs or librating in an MMR.
    Inspecting the Category 3 TTVs of the $N$-body-generated sample, we find that significant residuals are most frequently caused by these resonant effects.
}

\begin{table}[]
    \centering
    \begin{tabular}{| c | c c c c | c|}
     & \multicolumn{4}{|c|}{$N$ transiting} & \\ 
 $N$ planets & 1 &	2 &	3 &	4& Total \\ \hline
1 &1965  & 0  & 0  & 0  & 1965  \\ 
2 &1055  & 73  & 0  & 0  & 1128  \\ 
3 &374  & 50  & 2  & 0  & 426  \\ 
4 &141  & 26  & 1  & 0  & 168  \\ 
5 &46  & 11  & 3  & 0  & 60  \\ 
6 &24  & 6  & 1  & 0  & 31  \\ 
7 &7  & 0  & 0  & 0  & 7  \\ 
8 &2  & 0  & 0  & 0  & 2  \\ 
9 &2  & 0  & 0  & 0  & 2  \\ 
\hline
Total &3616  & 166  & 7  & 0  & 3789 \\

    \hline
    \end{tabular}
    \caption{Breakdown of the planet and transit multiplicities for low-$i$ planet population  (Compare with Table \ref{tab:multi_tranet_summary}). }
    \label{tab:multi_tranet_summary_low_i}
\end{table}

\begin{table}[]
    \centering
    \begin{tabular}{ c | c c c c |}
    	& Category 1: & Category 2:  & Category 3: & Singles \\  
	& Mass Measured & Significant, No Mass & Residuals &  \\  
 \emph{Mission}  & Med. (min., max.) & Med. (min., max.) & Med. (min., max.) & Med. (min., max.) \\ \hline 
Primary & 10 (6, 12) & 8 (4, 12) & 4 (2, 7) & 10 (8, 14)\\
$E_{3,NNN}$ & 16 (13, 18) & 15 (13, 19) & 9 (7, 10) & 37 (33, 40)\\
$E_{3,NSN}$ & 17 (12, 20) & 23 (20, 26) & 11 (9, 12) & 49 (43, 53)\\
$E_{4,NNN}$ & 19 (18, 21) & 14 (12, 18) & 9 (8, 11) & 43 (38, 49)\\
$E_{4,NSN}$ & 20 (18, 24) & 21 (16, 24) & 11 (9, 13) & 54 (49, 59)\\

    \hline
    \end{tabular}
    \caption{Summary of the expected TTV yields for different {\it TESS} mission scenarios for the low-$i$ planet population. (See Table \ref{tab:ttv_summary} for full description.)}
    \label{tab:ttv_summary_low_i}
\end{table}

\begin{table}[]
    \centering
    \begin{tabular}{ c | c c c c |}
    	& Category 1: & Category 2:  & Category 3: & Singles \\  
	& Mass Measured & Significant, No Mass & Residuals &  \\  
 \emph{Mission}  & Med. (min., max.) & Med. (min., max.) & Med. (min., max.) & Med. (min., max.) \\ \hline 
Primary & 6 (2, 9) & 7 (3, 13) & 13 (7, 18) & 40 (34, 49)\\
$E_{3,NNN}$ & 5 (2, 8) & 10 (6, 15) & 27 (19, 31) & 82 (74, 90)\\
$E_{3,NSN}$ & 6 (2, 10) & 12 (7, 16) & 30 (25, 35) & 95 (86, 101)\\
$E_{4,NNN}$ & 5 (2, 8) & 10 (7, 13) & 33 (29, 38) & 105 (98, 116)\\
$E_{4,NSN}$ & 6 (1, 9) & 15 (11, 19) & 37 (31, 43) & 122 (110, 129)\\

    \hline
    \end{tabular}
    \caption{Summary of the expected TTV yields for different {\it TESS} mission scenarios using TTVs synthesized via $N$-body simulations with TTVFast \citep{Deck14} rather than the analytic TTV model. (See Table \ref{tab:ttv_summary} for full description.)}
    \label{tab:ttv_summary_Nbody}
\end{table}

\section{Construction of TTV Basis Functions}
\label{SECN:APP:A}
Here we describe our method for computing the basis functions used in our linear TTV model.
In Equation \eqref{eq:TTV:LINEAR:linear_model} we approximate the $i$th planet's TTV induced by the $j$th planet at its $n$th transit as 
\begin{equation}
    \mu_j\left( \delta t^{(0)}_{i,j}(n)+  \text{Re}[{\Z}_{i,j}]t^{(1,x)}_{i,j}(n) + \text{Im}[{\Z}_{i,j}]t^{(1,y)}_{i,j}(n) \right)~.
\end{equation}
 We assume that the basis functions  $t^{(1,x)}_{i,j}(n)$ and $t^{(1,y)}_{i,j}(n)$ can be approximated by the contribution of the nearest first-order $p:p-1$ MMR and use the expressions derived in \citet{Lithwick12} and \citet{HL16}. Accordingly
\begin{eqnarray}
 t^{(1,x)}_{i,j}(n)  = P_i \times A\sin[2\pi t/P^\text{sup}+\phi]\\
 t^{(1,y)}_{i,j}(n)  = P_i \times A\cos[2\pi t/P^\text{sup}+\phi]
 \end{eqnarray}
 where the amplitude ($A$), phase ($\phi$), and super-period ($P^\text{sup}$) depend on the inner and outer planets' orbital periods, $P$ and $P'$, respectively, through  their fractional distance to resonance, $\Delta = \frac{(p-1)P}{pP'}-1$, and are given by
\begin{eqnarray}
 A &=& \begin{cases}
 \frac{1}{\pi}\frac{3(1-p)}{2p^2\alpha^2\Delta^2}\sqrt{f_{27}^2+f_{31}^2} & P_i<P_j \\
 \frac{1}{\pi}\frac{3}{2p\Delta^2}\sqrt{f_{27}^2+f_{31}^2} & P_i > P_j
 \end{cases}\\
 P^\text{sup}_j &=& \frac{P'}{p\Delta}\\
 \phi_j &=& p\lambda'_0 - (p-1)\lambda_0
\end{eqnarray}
where $\lambda_0$ and $\lambda'_0$ are the mean longitudes of the inner and outer planet, respectively, at $t=0$, $\alpha=a/a'$ is the semi-major axis ratio, and $f_{27}$ and $f_{31}$ are disturbing function coefficients, given in Appendix B of \citet{MDbook}, that depend on both $\alpha$  and the particular $p$:$p-1$ resonance considered.

We present a novel method for computing $t^{(0)}_{i,j}(n)$, the zeroth-order component (in the planets' eccentricities) of the TTV, that does not require us to truncate an infinite sum of Fourier terms unlike past perturbative TTV solutions \citep[e.g.,][]{Agol05,Nes09,DA15,HL16} or construct basis functions via $N$-body integrations \citep{Linial2018}. Our notation and derivation closely follows \citet{HL16}, where additional details on deriving analytic TTV expressions can be found.  We will begin by considering the TTVs of a transiting planet with orbital period $P$ subject to an external perturber with planet--star mass ratio $\mu'$ and orbital period $P'$. The case of an outer planet subject to an interior perturber is treated in Appendix \ref{sec:app:outer}. Hereafter, primed and unprimed quantities refer to the exterior and interior planet, respectively.  
\subsection{Inner Planet}
We write the time dependence of the planet's complex eccentricity, $z$, and mean longitude, $\lambda$, of the planet as
\begin{eqnarray}
    z(t) &=& z_0 + \delta z(t)\\
    \lambda(t) &=&\frac{2\pi}{P}(t-T) + \delta \lambda(t)    
\end{eqnarray}
where $\delta z(t)$ and $\delta \lambda(t)$ represent the deviations from a purely Keplerian orbit induced by the perturber. 
The planet's TTV is then given as a function of time by
\begin{equation}
    \delta t(t) =-\frac{P}{2\pi}\left[\delta \lambda(t) +i\delta z(t) -i \delta z^*(t) \right]+{\cal O}(e^2)~.\label{eq:thettv}
\end{equation}
The equations of motion are
\begin{eqnarray}
\frac{d\ln a}{dt}&=&2n'\frac{\mu'}{\sqrt{\alpha}}\pd{R}{\lambda}\label{eq:adot}\\
\frac{d\delta\lambda}{dt}&=&-\frac{3n'}{2\alpha^{3/2}}\frac{\delta a}{a}-2n'\mu'\sqrt{\alpha}\pd{R}{\alpha}\label{eq:lambdadot}\\
\frac{dz}{dt}&=&2in'\frac{\mu'}{\sqrt{\alpha}}\pd{R}{z^*}\label{eq:zdot}
\end{eqnarray}
where $\alpha=a/a'$ and $R$ is the disturbing function,
\begin{equation}
    R=a'\left(\frac{1}{|r'-r|}-\frac{r\cdot r'}{|r'|^3}\right).
\end{equation}
The disturbing function expanded to first order in eccentricity can be written
\begin{eqnarray}
R= \frac{1}{\sqrt{1+\alpha^2-2\alpha\cos\psi}}-\alpha\cos\psi+\left(z^*\frac{\partial R}{\partial z^*}\bigg|_{z,z^*=0}+z\frac{\partial R}{\partial z}\bigg|_{z,z^*=0}\right) + {\cal O}(e^2)
\label{eq:Rexpand}
\end{eqnarray}
where 
\begin{eqnarray}
    \frac{\partial R}{\partial z^*}\bigg|_{z,z^*=0} =\left(
        \frac{\alpha^2+\alpha\cos\psi+2i\alpha\sin\psi}{\left(1+\alpha^2-2\alpha\cos\psi\right)^{3/2}}+\frac{\alpha}{2}\cos\psi+i\alpha\sin\psi
    \right)\exp\left[{i\lambda}\right]
\end{eqnarray}
and $\psi=\lambda'-\lambda$.

We begin by solving for $\delta \lambda(t)$ to zeroth order in eccentricity.  Let us define $\Delta n = n'-n$ so that, ignoring the effect of planet-planet interactions, $\psi=\Delta n t+\psi_0$, where $\psi_0$ is determined by initial conditions. Inserting Equation \eqref{eq:Rexpand} into Equation \eqref{eq:adot}, to zeroth order in eccentricity and first order in $\mu'$ we have 
\begin{equation}
    \frac{\delta a(t)}{a}  = \frac{2n'\mu'}{\sqrt{\alpha}\Delta n}\osc{\frac{1}{\sqrt{1+\alpha^2-2\alpha\cos(\Delta n t+\psi_0)}}-\alpha\cos(\Delta n t+\psi_0)}
    \label{eq:a-soln}
\end{equation}
where the braces indicate the oscillating part of the enclosed expression, i.e.,
\begin{eqnarray}
    \osc{f(t)}=f(t)-\frac{\Delta n}{2\pi}\int_{0}^{2\pi/\Delta n}f(t')dt'~.
\end{eqnarray}
Inserting Equation \eqref{eq:a-soln} into Equation \eqref{eq:lambdadot} we derive 
\begin{equation}
\delta\lambda (t)= \mu'\osc{
\fracbrac{n'}{\Delta n}^2A(\Delta n t + \psi_0) + \frac{n'}{\Delta n}B(\Delta n t + \psi_0) }
\end{equation}
where
\begin{eqnarray}
A(x) &=& \frac{3}{\alpha^{2}}\left[
    \frac{2}{(1-\alpha)}K\left(\frac{x}{2}
\bigg|m\right)
-\alpha\sin\psi \right]\\
B(x) &=&-\frac{2\sqrt{\alpha}}{1-\alpha^2}\left[
        \left(1-\alpha^{-1}\right)E\left(\frac{x}{2}\bigg|m\right)
        +(1+\alpha^{-1})K\left(\frac{x}{2}\bigg|m\right)
        \right. \nonumber\\
&&+\left.\left(1-\alpha^2-\frac{2}{\sqrt{1+\alpha^2-2\alpha\cos\psi}}\right)\sin\psi\right],
\end{eqnarray}
, $K$ and $E$ are incomplete elliptic integrals of the first and second kind, respectively, and  $m=-\frac{4\alpha}{(1-\alpha)^2}$.
Note that the oscillating parts of elliptic functions are given by 
\begin{eqnarray}
    \osc{K(x|m)} &=&{K}(x|m)-\frac{2x}{\pi}{K}(m)\\
    \osc{E(x|m)} &=&{E}(x|m)-\frac{2x}{\pi}{E}(m)
\end{eqnarray}
where ${K}(m)$ and $E(m)$ are complete elliptic integrals of the first and second kind, respectively. \citet{Agol05} derive a similar closed-form expression for $\delta \lambda$ in their TTV formulas \citep[see also][]{Malhotra1993}.

Inserting Equation \eqref{eq:Rexpand} into Equation \eqref{eq:zdot}, the equation of motion for $z(t)$ to zeroth order in eccentricity is given by
\begin{eqnarray}
    \frac{dz}{dt} = 2in'\frac{\mu'}{\sqrt{\alpha}}\left(\frac{\alpha^2+\alpha\cos\psi+2i\alpha\sin\psi}{\left(1+\alpha^2-2\alpha\cos\psi\right)^{3/2}}+\frac{\alpha}{2}\cos\psi+i\alpha\sin\psi\right)\exp\left[{i\lambda}\right]~.\label{eq:zdot0th}
\end{eqnarray}
Unlike $\delta\lambda(t)$, Equation \eqref{eq:zdot0th} cannot be integrated to produce a simple closed-form solution for $\delta z(t)$. Rather than relying on a Fourier expansion of Equation \eqref{eq:zdot0th}, we simply integrate Equation \eqref{eq:zdot0th}  numerically after inserting $\psi=\Delta n t + \psi_0$ and $\lambda = nt +\lambda_0$ to obtain $\delta z(t)$ to zeroth order in eccentricity and first order in $\mu'$. Combining our numerical solution to Equation \eqref{eq:zdot0th} with Equation \eqref{eq:lambdadot}, we obtain the $t^{(0)}_{i,j}$ via Equation \eqref{eq:thettv}.

\subsection{Outer Planet}
\label{sec:app:outer}
Here we provide the TTV of a planet perturbed by an interior companion. The derivation is essentially the same as above so we do not reproduce all the steps but instead mention some minor modifications and quote the final results. First, for an outer planet the disturbing function is given by 
\begin{equation}
    R'=a'\left(\frac{1}{|r'-r|}-\frac{r\cdot r'}{|r|^3}\right)~.\label{eq:Rprime}
\end{equation}
and the equations of motion are 
\begin{eqnarray}
\frac{d\ln a'}{dt}&=&2n'\mu\pd{R'}{\lambda'}\label{eq:aPrimedot}\\
\frac{d\delta\lambda}{dt}&=&-\frac{3n'}{2}\frac{\delta a'}{a'}-2n'\mu\left(1+\pd{}{\alpha}\right){R'}\label{eq:lambdaPrimedot}\\
\frac{dz'}{dt}&=&2in'\mu\pd{R'}{z'^*}\label{eq:zPrimedot}
\end{eqnarray}

The outer planet's disturbing function, Equation \eqref{eq:Rprime}, expanded to first order in $z'$ becomes
\begin{eqnarray*}
    R'&=&\left(\left(1+\alpha^2-2\alpha\cos\psi\right)^{-1/2}-\alpha^{-2}\cos\psi\right)\\
&+&\left(\frac{\alpha\cos\psi-1+2i\alpha\sin\psi}{2\left(1+\alpha^2-2\alpha\cos\psi\right)^{3/2}}+\frac{1}{2\alpha^2}\left(\cos\psi-2i\sin\psi\right)\right)z'^*e^{i\lambda'} + c.c.~.
\end{eqnarray*}
The solution for $\delta \lambda'$ is given by
\begin{equation}
\delta\lambda' (t)= -\mu\osc{
\fracbrac{n'}{\Delta n}^2A'(\Delta n t + \psi_0) + \frac{n'}{\Delta n}B'(\Delta n t + \psi_0) }
\end{equation}
where
\begin{eqnarray}
    A'(\psi) &=& -\alpha^2 A(\psi)+3 (\alpha^{-2}-\alpha)\sin\psi~\\
    B'(\psi) &=& -\sqrt{\alpha}B(\psi)-\frac{2\alpha^2}{3}A(\psi)-(4\alpha+2\alpha^{-2})\sin\psi
\end{eqnarray}
and the solution for  $\delta z'(t)$ is obtained by numerically integrating
\begin{eqnarray}
    \frac{dz'}{dt} = 2in'\mu\left(\frac{\alpha\cos\psi-1+2i\alpha\sin\psi}{2\left(1+\alpha^2-2\alpha\cos\psi\right)^{3/2}}+\frac{1}{2\alpha^2}\left(\cos\psi-2i\sin\psi\right)\right)\exp\left[{i\lambda'}\right]~.\label{eq:z1dot0th}
\end{eqnarray}
with respect to time after substituting $\psi=\Delta n t+\psi_0$ and $\lambda'=n't+\lambda'_0$.

\bibliography{references}

\end{document}